\documentclass[11pt,a4paper]{article}

\usepackage[pdftex]{graphicx}
\graphicspath{{./figs/}}
\usepackage{hyperref}
\usepackage{url}
\usepackage[numbers,square]{natbib}
\usepackage{amsmath}    
\usepackage{amsfonts} 
\usepackage{amssymb}
\usepackage{caption}
\newenvironment{sciabstract}{%
\begin{quote}}
{\end{quote}}

\usepackage{authblk}

\begin{document}


\title{\textit{CompEngine}: a self-organizing, living library of time-series data}
\author[1,$\star$]{Ben D. Fulcher}
\author[2]{Carl H. Lubba}
\author[2]{Sarab S. Sethi}
\author[2,$\star$]{Nick S. Jones}
\affil[1]{School of Physics, The University of Sydney, Sydney, NSW, 2006, Australia}
\affil[2]{Mathematics Department, Imperial College London, Huxley Building, Queen’s Gate, London SW7 2AZ, UK}

\date{\today}
\maketitle

\begin{sciabstract}
\textbf{Summary:}
Modern biomedical applications often involve time-series data, from high-throughput phenotyping of model organisms, through to individual disease diagnosis and treatment using biomedical data streams.
Data and tools for time-series analysis are developed and applied across the sciences and in industry, but meaningful cross-disciplinary interactions are limited by the challenge of identifying fruitful connections.
Here we introduce the web platform, CompEngine, a self-organizing, living library of time-series data that lowers the barrier to forming meaningful interdisciplinary connections between time series.
Using a canonical feature-based representation, CompEngine places all time series in a common space, regardless of their origin, allowing users to upload their data and immediately explore interdisciplinary connections to other data with similar properties, and be alerted when similar data is uploaded in the future.
In contrast to conventional databases, which are organized by assigned metadata, CompEngine incentivizes data sharing by automatically connecting experimental and theoretical scientists across disciplines based on the empirical structure of their data.
CompEngine's growing library of interdisciplinary time-series data also facilitates comprehensively characterization of algorithm performance across diverse types of data, and can be used to empirically motivate the development of new time-series analysis algorithms.
\\
\textbf{Availability:} The website contains an initial set of over 24\,000 interdisciplinary time series, and can be accessed at \url{https://www.comp-engine.org/}.\\
\textbf{Contact:} \href{nick.jones@imperial.ac.uk}{nick.jones@imperial.ac.uk}, \href{ben.fulcher@sydney.edu.au}{ben.fulcher@sydney.edu.au}.
\end{sciabstract}

\section{Introduction}
Taking repeated measurements of a system over time, yielding time-series data, is ubiquitous in biology and medicine.
Biological systems vary across a wide range of spatial and temporal scales, from the microscopic spatial and fast temporal scales of molecular dynamics, through to the slow macroscopic dynamics of ecological populations.
Biological applications of time-series analysis are correspondingly diverse: from high-throughput phenotyping of model organisms---which involves vast quantities of measured time-series data---to the goals of precision medicine, in which information is extracted from diverse data streams including movement dynamics (during sleep), heart rates, and speech recordings.
Beyond biology and medicine, time series are studied in mathematics, statistics, and physics, and measured in disciplines ranging from economics to astrophysics and meteorology.
Unprecedented volumes of time series are also collected for diverse commercial applications, including fault identification from sensor recordings of industrial processes, fraud detection from vast streams of credit-card transactions, and marketing strategy development from the dynamics of online behaviors.
The wide range of problems involving time-series data has resulted in a diversity of analysis methods, but despite commonalities between time-varying systems studied in different contexts, time series and their methods are rarely compared \cite{Fulcher:2013ft}.

There is much to be gained from interdisciplinary collaboration on time-varying systems.
For example, connecting people studying similar real-world systems could prompt them to work collaboratively to understand common patterns in their data.
Similarly, connecting simulations of time-varying model systems (for which underlying mechanisms are known) to the types of real-world systems that exhibit similar dynamics could connect theoreticians and experimentalists to better understand the mechanisms underlying empirically observed statistical patterns.
Connecting real-world dynamics to model simulations through their properties is feasible \cite{Fulcher:2013ft}, having connected share-price dynamics to SDE models used in financial modeling, rainfall dynamics to Weibull-distributed noise processes (which are commonly used to model and predict extreme events in hydrology), and speech phonemes to oscillatory dynamical systems.
However, due to large barriers to identifying commonalities, and hence areas for productive collaboration around common problems, initiating meaningful interdisciplinary connections remains challenging.

Here we introduce CompEngine, a self-organizing library of interdisciplinary time-series data that automatically highlights meaningful interdisciplinary connections between empirical data.
CompEngine uses a common feature-based representation of time series to organize them according to the properties of their empirical dynamics.
CompEngine contains an initial set of over 24\,000 time series encompassing recordings from a wide range of:
(i) biological and medical systems, including birdsong, population dynamics, electrocardiograms (ECG), heart-rate intervals, gait, and tremor series;
(ii) non-biological systems, including audio, finance, meteorological, and astrophysical data;
and (iii) synthetic model systems, including data generated from simulating sets of differential equations, iterative maps, and stochastic processes \cite{Fulcher:2013ft}.
Each time series is characterized by user-provided metadata about what system was measured and how it was recorded.

As users upload and share their own data, connections between different data objects are updated automatically as the library grows and reorganizes itself.
This process often yields surprising interdisciplinary connections between the properties of empirical data generated from real-world systems and synthetic model-generated data, thus lowering the barrier for fruitful interdisciplinary collaboration by automatically connecting scientists through the data they analyze.
CompEngine's large interdisciplinary library of time-series data can be downloaded and used to calibrate the performance of new time-series analysis algorithms on diverse empirical data.
In this paper, we introduce CompEngine, explaining its motivation, the research underlying its functional machinery, and how we envisage its key functionality being of broad utility for time-series analysis.


\begin{figure*}[ht]
  \centering
    \includegraphics[width=.85\textwidth]{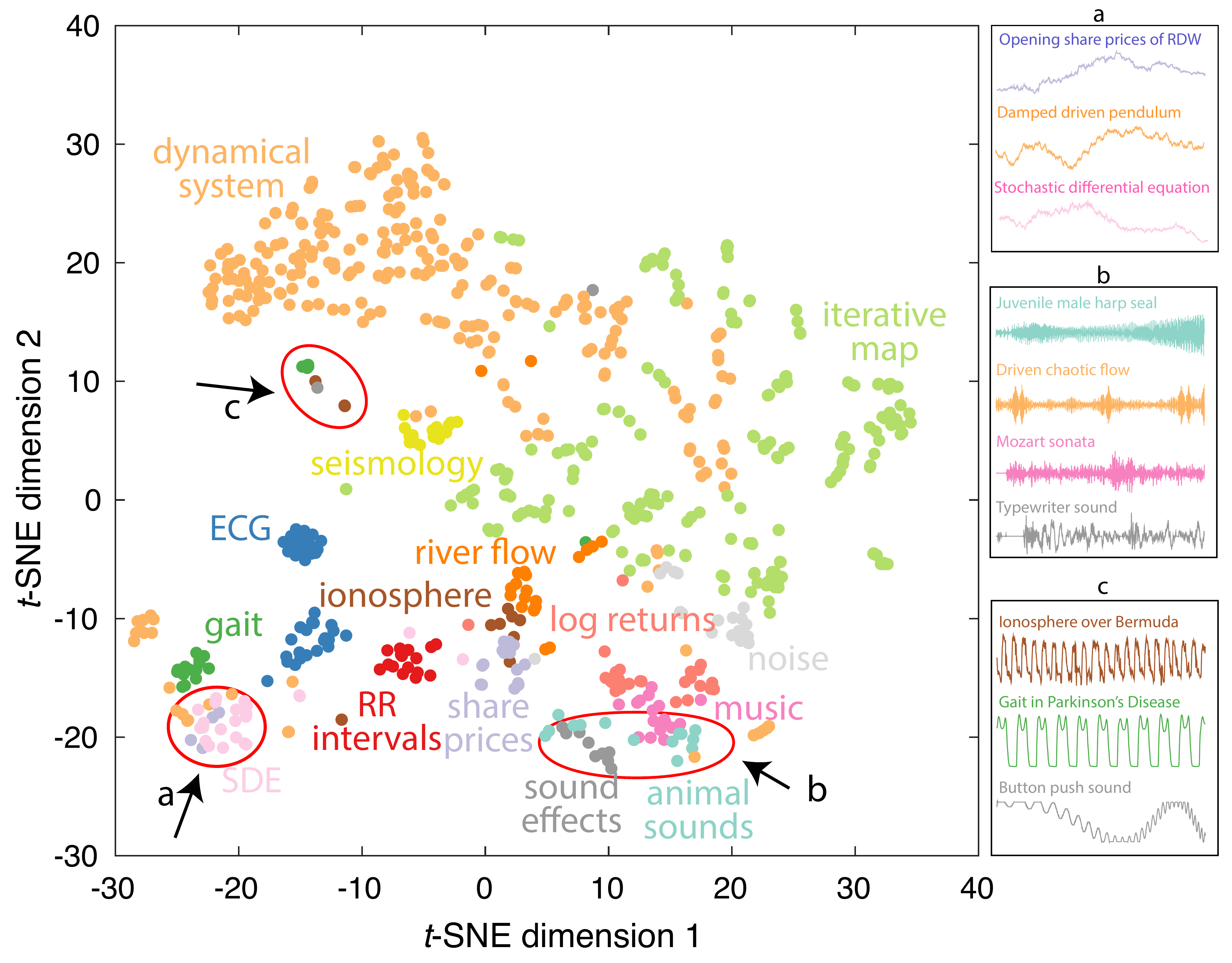}
  \caption{
  \textbf{Diverse time series, of varying types, sampling rates, and durations, are organized meaningfully in a common feature space.}
We plot each time series in a reduced, two-dimensional feature space, computed with $t$-SNE \cite{maaten2008visualizing}.
Each time series is labeled by color according to a set of broad categories (see Supplementary Information for descriptions of each category).
Most categories occupy distinct parts of the space, with mixing between similar categories according with intuition.
Interesting connections between distinct systems are flagged where distinct classes of data overlap; some examples of time series (visualized as 1000-sample segments) from three such areas of the space are shown in the right panel of the figure.
  }
  \label{fig:LowDim}
\end{figure*}

\section{Approach}
For a self-organizing library to meaningfully structure diverse data, it requires an appropriate measure of similarity between pairs of objects in the library.
The fundamental data object in CompEngine is a univariate and uniformly-sampled time series (or, generally, any data that can be represented as an ordered vector of real numbers \cite{Fulcher:2018vj}).
Our challenge is therefore to develop a similarity measure that can compare time series measured at different sampling rates, from different systems, and for different durations of time.
Drawing on previous research, we achieve this using a feature-based representation \cite{Fulcher:2013ft, Fulcher:2017fk, Lubba:2018}.

\subsection{Feature-based representations of time series}
Of the myriad ways in which two time series can be compared, measuring a set of properties of the measured dynamics allows a time series to be represented as a point in a common feature space, regardless of how/where it was measured \cite{Fulcher:2018vj}.
Such `feature-based' representations of time series have been used to successfully tackle a wide range of problems involving time series \cite{Fulcher:2018vj}, including classification (or regression) \cite{Fulcher:2013ft, Fulcher:2014uo, Fulcher:2017fk}, clustering \cite{wang2006}, forecasting \cite{Kang:2017hs}, and anomaly detection \cite{hyndman2015large}.

To generate a feature vector, a univariate time series of $T$ ordered measurements, $x_t$ ($t=1,2,...,T$), is mapped to a set of $F$ features, $f_i$ ($i = 1,2,...,F$), where each feature is the real-valued output of some algorithm applied to the time series.
The feature-based distance between two time series, $\mathbf{x}^{(j)}$ and $\mathbf{x}^{(k)}$, is then defined in terms of the distance between the feature vectors, $\mathbf{f}^{(j)}$ and $\mathbf{f}^{(k)}$, as $d\left[\mathbf{f}^{(j)},\mathbf{f}^{(k)}\right]$.
To weight all features equally in the distance calculation, the values of each feature are normalized across the time-series dataset; here we use an outlier-robust sigmoidal normalization \cite{Fulcher:2013ft}.
Our core problem then becomes how to define an $F$-dimensional feature space: $\mathbf{x} \rightarrow \mathbf{f} \in \mathbb{R}^F$ in which feature-based distances capture meaningful differences between the vastly different types of empirical dynamics contained in a general time-series data library like CompEngine.

\subsection{Using feature-based similarity to structure diverse data}

Feature-based representations provide a feasible approach through which diverse time series can be meaningfully compared; but how do we select a feature space in which to represent them, given the vast search space of interdisciplinary time-series analysis methods \cite{Fulcher:2018vj}?
Recent work introduced an approach known as `highly comparative time-series analysis' \cite{Fulcher:2013ft, Fulcher:2014uo, Fulcher:2018vj}, implemented in the software package, \textit{hctsa} \cite{Fulcher:2017fk}, which includes algorithmic implementations of over 7000 features from the time-series analysis literature.
To investigate whether a feature-based representation can organize different types of dynamics, we used \emph{hctsa} to generate feature-based representations of data from fifteen different classes, encompassing both:
\emph{simulated model systems}: deterministic dynamical systems, discrete iterative maps, stochastic differential equations, and random noise series;
and \emph{empirical data} from: seismology, river flow, share prices, log returns of financial data, ionosophere, sound effects, animal sounds, music, electrocardiograms, heart-rate (RR) intervals, and gait dynamics.

As shown in Fig.~\ref{fig:LowDim}, a reduced, two-dimensional $t$-SNE projection of the full feature space \cite{maaten2008tSNE}, yields a meaningful representation of the dataset, in which distinct categories of data occupy characteristic parts of the space.
For example, the lower-right part of the space, labeled `b', contains audio data: clusters of sound effects, animal sounds, and music.
As shown in Fig.~\ref{fig:LowDim}b, time series in this region of the feature space are visually similar, including male juvenile harp seal audio, a downsampled excerpt from a Mozart sonata, the sound of a typewriter, and a numerically simulated chaotic flow that captures many of the oscillatory and bursty properties of real audio data ($\ddot{x} = -x^3 + \sin(\Omega t)$, $\Omega = 1.88$ \cite{sprott2003chaos}).
Other examples are annotated, including the slow fluctuations of time series in region `a', which contains opening share prices, a simulated damped driven pendulum, and the output from a stochastic differential equation (SDE); note that SDE models are frequently used to model financial data \cite{Fulcher:2013ft}.
Periodic dynamics are concentrated in region `c', including gait dynamics of patients with Parkinson's disease, ionosphere measurements, and the audio of a button-push sound effect.
Overlaps between the structure of real-world dynamics and that of synthetic time-varying systems---such as the connection highlighted here between share price fluctuations and those of useful SDE models---highlights fruitful possible connections between theory and experiment \cite{Fulcher:2013ft}.
This ability to make meaningful connections between diverse types of time-series data through a common feature space representation forms the basis for CompEngine being an informative self-organizing library of time-series data.

\begin{figure}[h]
  \centering
    \includegraphics[width=.50\textwidth]{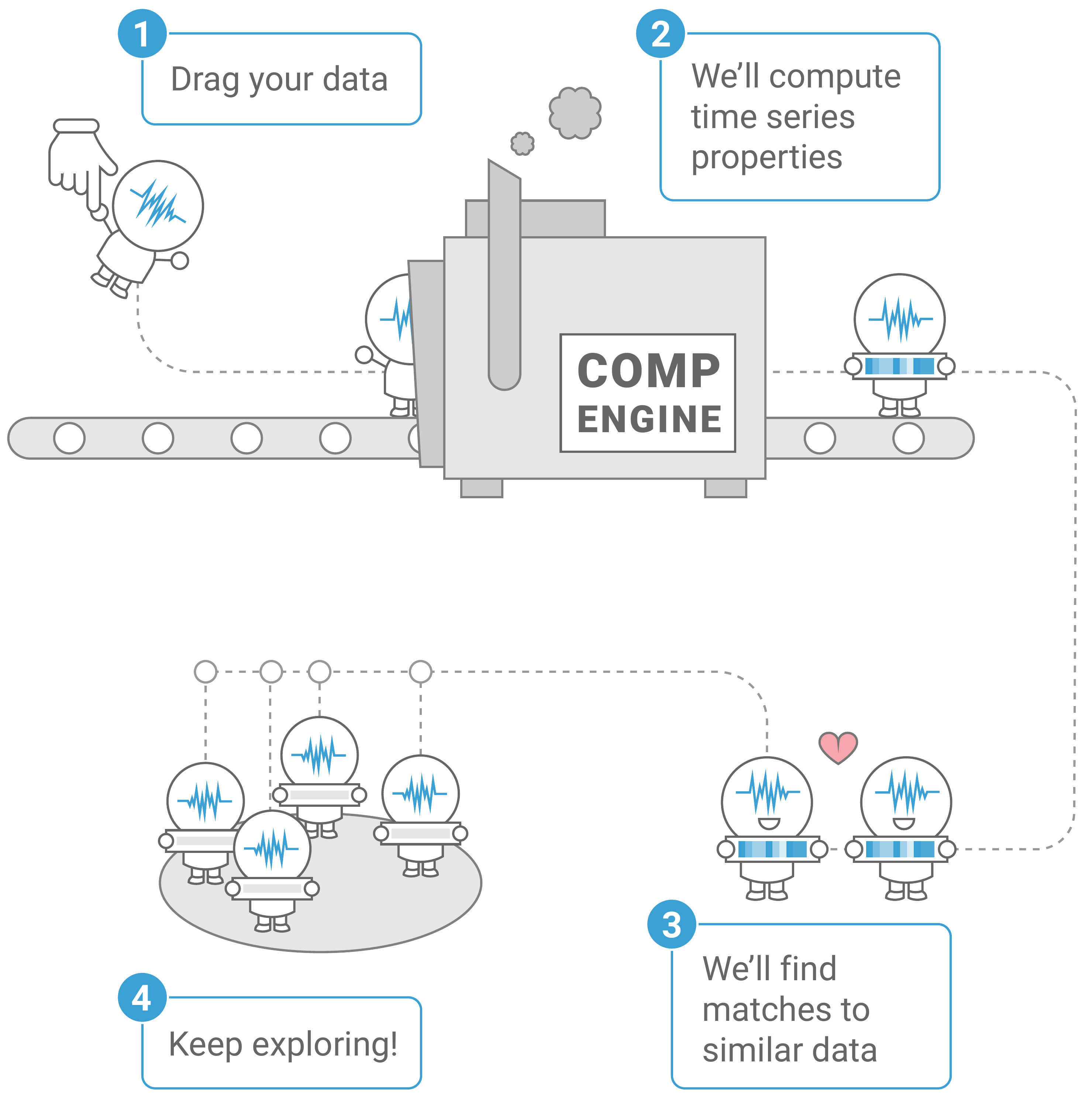}
  \caption{
  \textbf{Using CompEngine to understand interdisciplinary connections between time series}.
    After uploading a time series (Step 1, Sec.~\ref{sec:upload}), we compute a set of features (Step 2, Sec.~\ref{sec:canonicalFeatureVector}) which are used to calculate a similarity score between the new data and all existing data in the library (Step 3).
    The data context can then be explored through an interactive visualization (Sec.~\ref{sec:interactiveVis}), and the data can be contributed to grow our time-series library.
  }
  \label{fig:pipelineSchematic}
\end{figure}

\subsection{A canonical time-series feature vector}
\label{sec:canonicalFeatureVector}
Using \emph{hctsa} to convert time series to comprehensive feature vectors requires the computationally expensive step of calculating over 7000 features.
In recent work, this large feature set was reduced to a smaller subset of interpretable features that show high classification performance and exhibit minimal redundancy across a wide range of classification tasks, yielding a set of 22 features called \emph{catch22} \cite{Lubba:2018}.
These features are implemented efficiently in C and capture the conceptual diversity present in \emph{hctsa}, incorporating measures of autocorrelation, predictability, stationarity, distribution of values, and self-affine scaling \cite{Lubba:2018}.
Compared to the full \emph{hctsa} feature set, distances between pairs of time series are highly correlated in the space of \emph{catch22}, $r = 0.77$ (using the time series analyzed above, cf. Fig.~\ref{fig:LowDim}).
Thus, \emph{catch22} structures time-series datasets in a similar way to the full \emph{hctsa} feature set but computes without a software license and at a small fraction of the computational cost, making \emph{catch22} a much more scalable solution for a resource like CompEngine.

\section{Functionality}

\subsection{Retrieving an interdisciplinary context for time series}
Representing time series in a common feature space, implemented as \emph{catch22} \cite{Lubba:2018}, allows us to construct a self-organizing time-series library: CompEngine.
Figure~\ref{fig:pipelineSchematic} illustrates the basic pipeline through which a user can upload their time series and visualize connections to similar data in the CompEngine library.

\begin{figure*}[ht!]
  \centering
    \includegraphics[width=.90\textwidth]{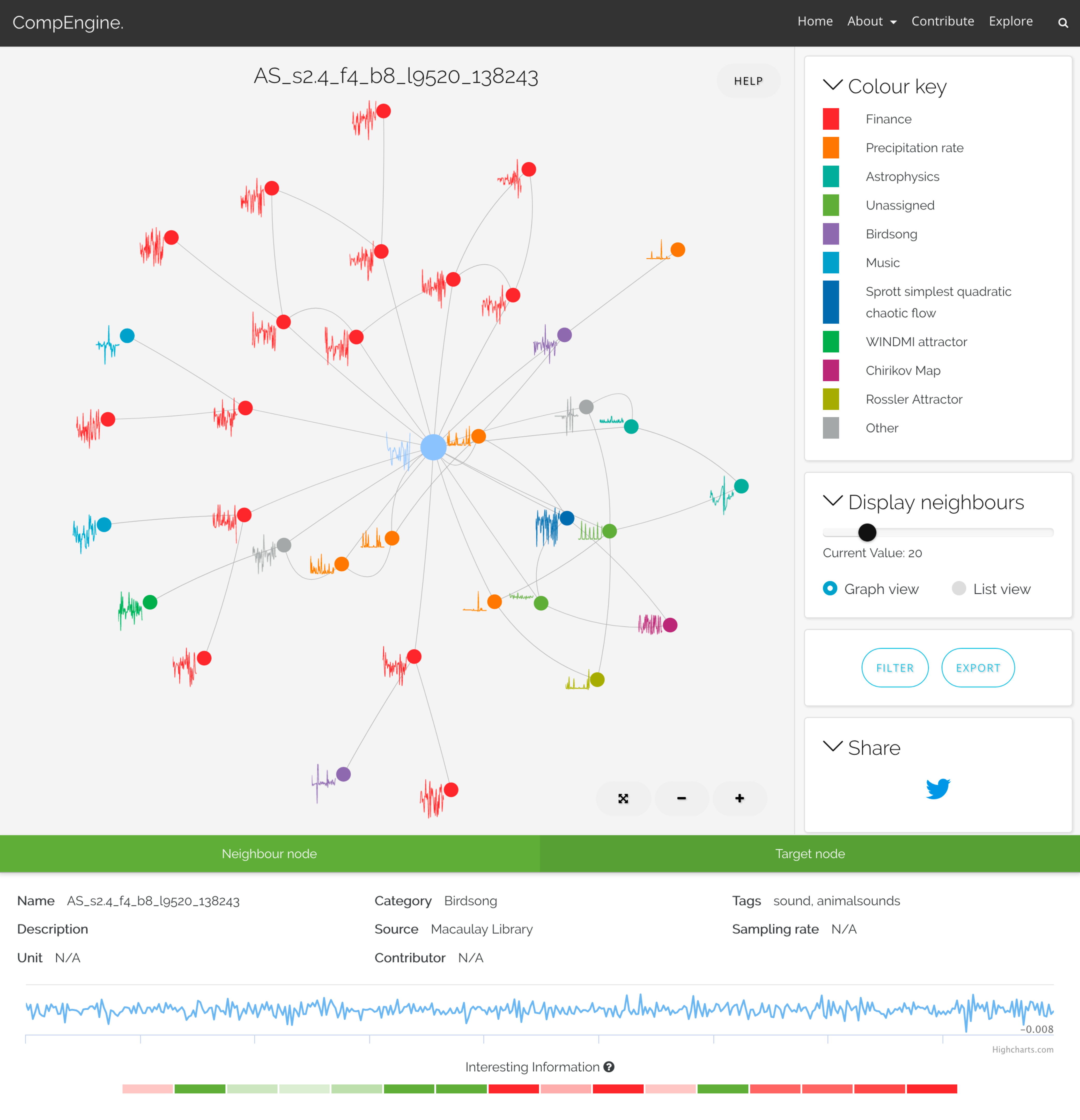}
  \caption{
\textbf{\emph{CompEngine} is a web-based platform that allows users to upload their time-series data, and automatically find connections to other types of data in our library}.
    Here we show CompEngine's web-based interface for visualizing similar types of time-series data as an interactive network.
    The target time series---in this example birdsong from the black-bellied plover---is shown as the central light blue node, with neighbors plotted around it, colored according to their categorical assignment (labeled in the `Colour key').
    Neighbors range from other types of audio, e.g., birdsong (purple) and music (blue), as well as data generated from time-series models, e.g., WINDMI attractor (forest green) and the Chirikov Map (violet).
    Information about the target node is shown in the lower panel, including metadata, an interactive visualization of the time series, and a set of computed feature values.
    These feature values are visualized using a traffic-light colour scheme, highlighting time-series properties that are exceptionally high (green) or low (red) compared to the rest of the CompEngine library (information is shown on hover).
  }
  \label{fig:CompEngine}
\end{figure*}

\subsubsection{Data upload} \label{sec:upload}
Upload supports text files (\verb|.txt|, \verb|.csv|) and Excel files (\verb|.xls|, \verb|.xlsx|) containing a single column vector of real numbers.
Audio data upload is also supported (\verb|.mp3| or \verb|.wav|; the audio is converted to a time series using the first audio channel with floating-point encoding and a minimum sampling rate of 4\,kHz).
For computational efficiency, individual time series containing more than 10\,000 samples are truncated to the first 10\,000 samples.
CompEngine also supports upload of multiple univariate time series through a bulk upload function.
All uploaded data is licensed under the `no rights reserved' CC0 license.

Uploaded data can be permanently added to the CompEngine library by providing basic metadata that is sufficient to allow new users to be able to understand it.
This includes:
\textit{Name}, \textit{Sampling Rate}, and \textit{Description}, as well as:
\textit{Source} (identifies who/where/how the data were measured),
\textit{Category} (a hierarchical categorization of time-varying synthetic and real-world systems), and
\textit{Tags} (any additional keywords).
To facilitate automatic connections to new data, users can optionally choose to provide contact information and opt-in to regular updates when new data is uploaded to CompEngine in the future that is similar to the uploaded time series.

\subsubsection{Interactive visualization of similar time series} \label{sec:interactiveVis}
As shown in Fig.~\ref{fig:CompEngine}, CompEngine visualizes the nearest neighbors to a target time series as an interactive network.
Each node corresponds to a time series, and is coloured by its category label and annotated with a representative time trace.
Links in the network capture the feature-vector similarity between pairs of time series.
Users can interactively zoom in or out of the network, filter on the categories of retrieved neighbors, export all data (as a \verb|.json| or zipped \verb|.csv| file), and share the results (including as a URL).

Detailed inspection of time traces and basic metadata of neighboring nodes can be done using the inspector panel (at the bottom of the interface).
Double-clicking on a neighboring time series allows the user to go to the neighborhood of that time series, allowing progressive exploration of the database through local neighborhoods.
Users can also access basic information about the canonical feature vector of their data to understand where their data is placed with respect to other data in the library for each feature, flagging cases where the data is exceptional, e.g., `the lag-9 autocorrelation of this time series is in the top 1\% of data in our library'.

\section{Utility}
In the context of time-series analysis, CompEngine makes a large library of time-series data freely available to the public that will grow over time through community contributions and can be explored by both metadata and dynamical properties.
Two examples of new types of science that is facilitated by CompEngine are described below.


\subsection{Connecting scientists through their data}
A self-organizing library structures data objects empirically according to their properties, not their assigned metadata.
As described above, and shown in Fig.~\ref{fig:LowDim}, this can flag unexpected connections, including:
(i) between real-world and simulated data---suggests relevant mechanistic or statistical models for a real-world system; and
(ii) between empirical dynamics of two real-world systems---highlights new opportunities to collaborate across interdisciplinary borders to understand connections between seemingly disparate systems.
As well as being provided interactively at the time of upload, CompEngine also continues to search for new matches as additional data is uploaded in the future, alerting the user by email to future matches as they occur.
By treating theoretical and diverse types of empirical time series in the same way, CompEngine thus provides a direct incentive to data sharing: the user learns more about how their system of interest relates to other synthetic and real-world systems, both immediately at the time of upload, and into the future as the time-series data library evolves.


\subsection{Diverse data for algorithm development}
In practice, the selection of a time-series algorithm for a given application is based on the subjective experience of the data analyst.
Moving towards a more systematic procedure requires a comprehensive understanding of the characteristics of the data that a given algorithm performs well on \cite{SmithMiles:2014kf}, given that no single algorithm can exhibit strong performance on all types of data \cite{wolpert1997no}.
CompEngine provides access to a large and growing data repository to facilitate evaluation of algorithms on diverse time-series data, allowing us to comprehensively and objectively understand the strengths and weaknesses of different time-series analysis algorithms applied to different types of data.
This process may indeed highlight unexpected examples of datasets for which the new method performs well on, inspiring new interdisciplinary collaborations on common problems.
CompEngine can thus be used to guide the presentation of new time-series analysis algorithms without data-selection bias.
This would allow us to understand the applicability of different algorithms to different types of datasets, such that future methods development could be empirically tailored to the types of problems that our current analysis toolbox performs poorly on.


\subsection{A template for other self-organizing data libraries}
CompEngine uses the example of time series to demonstrate how complex data objects can be projected in a common feature space and organized on the basis of their empirical properties.
The benefits of such a self-organizing library are applicable not just to the time-series analysis community, but also to other data objects, including complex networks \cite{agarwal2010high}, images, point clouds, and multivariate classification datasets \cite{munoz2018instance}.
We hope that CompEngine may serve as a template to modern self-organizing libraries of many such data types that encourage broader scientific collaboration on data.


%

\section{Summary}
While recent years have seen dramatic growth in data sharing, including in scientific research \cite{molloy2011open, gewin2016data}, data repositories are typically organized only on the basis of user-assigned metadata.
Here we introduce CompEngine, which adds a computational layer to self-organize a large repository of time-series data, automatically retrieving interesting connections between diverse time series.
To our knowledge, the platform is the first self-organizing collection of scientific data, containing an initial library of over 24\,000 diverse time series.
Compared to conventional data repositories, this provides a direct incentive for data sharing, with users immediately obtaining new understanding of the interdisciplinary context surrounding their data, and an option to be notified when similar data are uploaded in the future.
As well as facilitating connections between a large and diverse community studying time-series patterns in biological systems, this resource has a much broader reach: from individuals self-recording their heart-rhythms or sleep patterns through a wrist device, to those probing a portfolio of assets or examining drilling profiles.
We envisage CompEngine becoming a unifying portal that links disparate users---be they scientists or data analysts---who currently work isolated from one another due to high barriers to comparison, and hence collaboration.







\section*{Acknowledgements}
We thank Rachael Fulcher for designing the infographic in Fig.~\ref{fig:pipelineSchematic}.
Implementation of the backend algorithms for nearest-neighbor matching and general web and database design and implementation were done by Vokke Pty Ltd.

\section*{Funding}
NSJ would like to thank the EPSRC for grants EP/K503733/1 and EP/N014529/1 and NERC for the grant NE/K007270/1.
BDF is supported by an NHMRC Fellowship, 1089718.
CHL thanks the EPSRC for grant EP/L016737/1.

\bibliographystyle{benbibstyle}


\clearpage
\section*{Supplementary Information}

\subsection*{Low-dimensional projection of diverse data}
Our example of a low-dimensional projection of diverse time series represented as \emph{hctsa} feature vectors \cite{Fulcher:2017fk}.
Features were normalized according to a scaled sigmoidal transformation \cite{Fulcher:2013ft}, filtering out columns with any bad values, resulting in a data matrix of dimensions $786 \times 5131$ (time series $\times$ features).
The dimensionality reduction was computed using $t$-SNE \cite{maaten2008tSNE}, after reducing the data matrix to a space of 50 principal components, and using the barnes-hut approximation for $t$-SNE (implemented in Matlab as \texttt{tsne}).
Our example included a subset of data classes taken from a set of 1000 empirical time series \cite{Fulcher2017Empirical}.
We restricted the analysis to data labeled into the following 15 groups:
`dynamical system': 274 time series simulated numerically from diverse systems of ordinary differential equations (ODEs);
`iterative map': 221 time series simulated numerically from diverse iterative maps \cite{sprott2003chaos};
`SDE': 23 time series simulated numerically from stochastic differential equations;
`noise': 20 time series of uncorrelated random noise taken from distributions including beta, binomial, and normal;
`seismology': 16 time series taken from a dataset of earthquakes and explosions;
`ionosphere': 13 recordings from the Earth's ionosphere, taken from SPIDR \cite{zhizhin2008space};
`ECG': 40 ECG time series from Physionet (varying in length from 1200 to 10\,000 samples) \cite{moody2001physionet};
`gait': 20 gait time series, 19 of which are from Physionet \cite{moody2001physionet};
`RR intervals': 16 RR interval series (varying in length from 1900 to 8600 samples) \cite{moody2001physionet};
`music': 17 time series taken from downsampled audio recordings of music in styles ranging from Baroque through to post-metal;
`sound effects': 12 time series taken from audio recordings of various sound effects, including a doorbell sound, a food processor, and urination,;
`animal sounds': 14 time series taken from the Macauley Library (\url{https://www.macaulaylibrary.org/}), including sounds from the Harp Seal, Marbled Wood-Quail, and the Red Junglefowl;
`share prices': 19 share price time series from Yahoo Finance; and
`log returns': log-return transformations of financial time series.

%
%
%
%
%
%

\end{document}